\documentclass[aip,amsmath,amssymb,reprint]{revtex4-2}
\usepackage{epsfig}
\usepackage{wrapfig}
\usepackage{bbm}
\usepackage[usenames]{color}
\usepackage{array}
\usepackage{times}

\usepackage[normalem]{ulem}

\usepackage{graphicx}
\usepackage{physics}
\usepackage{svg}
\usepackage{float}
\usepackage{multirow}
\usepackage{natbib,twoopt}
\usepackage[breaklinks=true]{hyperref}
\usepackage{xcolor}
\usepackage{amsmath,amsfonts,amssymb}
\usepackage{graphicx}
\usepackage{hyperref}
\usepackage{color}
\usepackage{stackengine}
\usepackage{subfigure}
\usepackage{verbatim}
\usepackage{comment}
\usepackage{stmaryrd} 

\newcommand{\df}[2]{\frac{\partial #1}{\partial #2}}


\newcommand{\RNum}[1]{\uppercase\expandafter{\romannumeral #1\relax}}

\begin{document}

\title{Non-Hermitian excitations in nonlinear topological lattice}


\author{Vlad Simonian}
\affiliation{School of Physics and Engineering, ITMO University, Saint  Petersburg 197101, Russia}

\author{Daria A. Smirnova}
\affiliation{Research School of Physics, Australian National University, Canberra, ACT 2601, Australia}

\author{Maxim A. Gorlach}
\email{m.gorlach@metalab.ifmo.ru}
\affiliation{School of Physics and Engineering, ITMO University, Saint  Petersburg 197101, Russia}

\begin{abstract}
Non-linear effects and non-Hermitian phenomena unveil additional intricate facets in topological matter physics. They can naturally intertwine to enable advanced functionalities in topoelectrical circuits and photonic structures. 
Here, we illustrate the subtle 
interplay between nonlinearity and non-Hermiticity by examining the characteristics of small wave perturbations on 
the background of the self-induced topological edge state in the nonlinear Su-Schrieffer-Heeger model. We demonstrate that their underlying physics is captured by the non-Hermitian effective Hamiltonian, which features nonreciprocal coupling terms and entails unconventional time-dependent field localization.
\end{abstract}

\maketitle

\section{Introduction}

Topological states have attracted much attention offering extra resilience to disorder and imperfections originating from the global properties of the system. Being initially introduced in condensed matter context~\cite{Xiao2010,Hasan2010}, they were later generalized to many other wave phenomena including mechanics, acoustics, electric circuits and photonic systems~\cite{Ozawa_2019}.

One of the defining features of photonic systems is their open nature which leads to non-Hermitian effective Hamiltonians and associated non-Hermitian topological physics~\cite{Bergholtz_2021}. 
The representative toy model frequently employed to test the effects of non-Hermiticity on topology is a celebrated Su-Schrieffer-Heeger (SSH) model~\cite{Su1979,Heeger1988} [Fig.~\ref{nonlin}(a)]. Addition of non-Hermiticity to this one-dimensional (1D) system significantly alters its physics. One of such profound changes is the so-called non-Hermitian skin effect manifesting itself in the localization of all eigenstates at the edges of the system provided open boundary conditions are imposed~\cite{Helbig2020,Weidemann2020,Bergholtz_2021}. 

In particular, such behavior is observed when the couplings in the SSH model are made nonreciprocal such that the coupling of left site with the right is not equal by its magnitude to the coupling of right site with the left, as first introduced in Hatano-Nelson model~\cite{Hatano1997}. While being quite artificial, the hybrid of SSH and Hatano-Nelson models was actively explored theoretically and analytical solutions for bulk and edge modes were obtained exhibiting a number of distinctions from their Hermitian counterparts~\cite{Kunst_2018,Yao_2018}. For instance, while the conventional SSH model features only a single zero-energy mode at a given edge, its non-Hermitian generalization supports two modes localized at a given edge at zero frequency.

A parallel and seemingly independent line of research is presented by nonlinear topological photonics~\cite{Smirnova_2020}, which aims to harness nonlinearity readily available in many photonic systems to tailor topological phenomena. A promising possibility is to reconfigure the topological modes by changing the intensity of excitation. Such reconfigurability has been thoroughly explored for the SSH model with intensity-dependent couplings, first theoretically~\cite{Hadad_2016} and later experimentally in radiofrequency range by connecting $LC$ resonators via nonlinear varactor diodes~\cite{Hadad_2018}. A profound prediction in this area are self-induced topological states whose existence and the degree of localization are governed by the intensity of excitation. 

Despite these advances, the interplay of nonlinear and non-Hermitian phenomena 
remains barely understood, with only a few works charting this field~\cite{Poddubny_2023}. At the same time, the fusion of the two concepts is already opening fruitful applications such as topological lasers~\cite{Harari2018,Bandres2018,Dikopoltsev2021,Zhu2022}, making studies at the interface of these two areas a timely and significant problem.

In this Article, we aim to bridge this gap and explore the spectrum of perturbations of the nonlinear self-induced topological states in the SSH model with intensity-dependent couplings. As we prove, the physics of those excitations is captured by the non-Hermitian effective Hamiltonian with nonreciprocal couplings, thus uncovering an unexpected connection between nonlinear and non-Hermitian topological physics. In addition, this result also suggests a straightforward experimental implementation of Hatano-Nelson-type nonreciprocal couplings by perturbing the nonlinear system.

The rest of the Article is organized as follows. In Section II, we revisit the steady-state solutions of the nonlinear SSH model in the form of the self-induced nonlinear edge states, discussing their intensity dependence and methods for computing their profiles. In Section III, we derive a non-Hermitian effective Hamiltonian describing small perturbations on the background of the nonlinear edge state. We show that such perturbations do not destroy the nonlinear mode and exhibit an unusual behavior: their real and imaginary parts (as defined with respect to the background mode) oscillate differently. In Section IV we analyze the localization of eigenmodes of system and discuss the non-Hermitian skin effect. We conclude with Section V, which discusses and summarizes the results. 

\section{Self-induced nonlinear edge states}\label{sec:edgestates} 

\begin{figure}
  \includegraphics[width=0.7\linewidth]{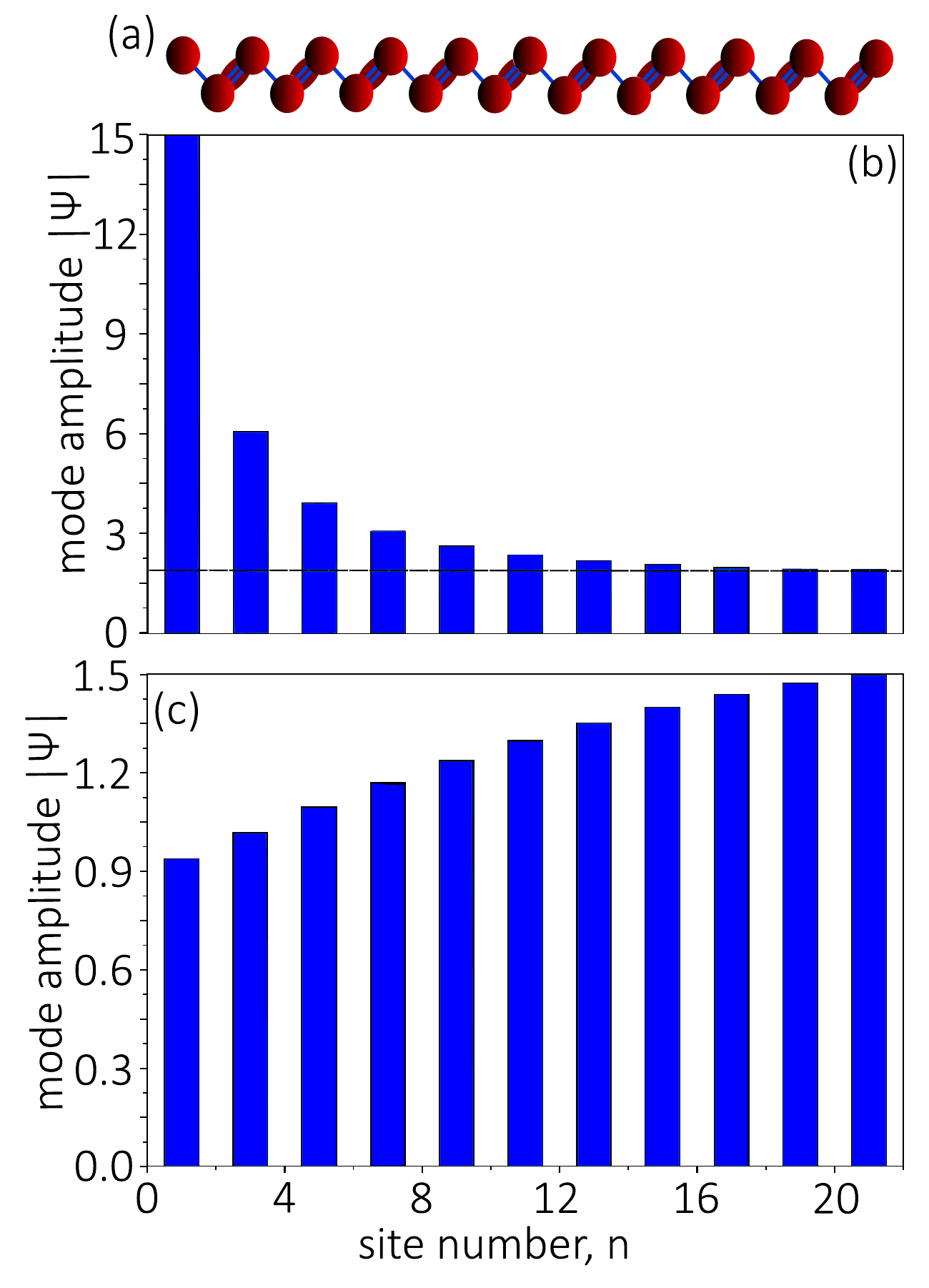}
    \caption{Nonlinear Su-Schrieffer-Heeger model and 
    steady-state solutions in a finite array of $27$ elements. (a) Schematic of the array. Intensity-dependent couplings are highlighted. (b,c) Profiles of the zero-frequency nonlinear topological edge mode in (b) high-intensity regime $I=15$; (c) low-intensity regime $I=1.6$. The rest of the parameters 
    are $\alpha = 5\times 10^{-5}, J_1 = 2.3\times 10^{-3},J_2 = 2\times 10^{-3}, I_{\text{cr}}  \approx 2.45 $ similar to Ref.~\cite{Hadad_2016} and consistent with experiments Ref.~\cite{Hadad_2018}. }
  \label{nonlin}
\end{figure}

We consider a nonlinear SSH model which is a 
1D array of single-mode cavities with the nearest-neighbor alternating couplings~\cite{Hadad_2016,Hadad_2018}, as depicted schematically in Fig.~\ref{nonlin}(a). In this model, the intracell coupling is linear, $J_1^{(2n-1,2n)}=J_1=\text{const}$, while the intercell coupling $J_2^{(2n,2n+1)}$, highlighted by red, is nonlinear and depends on the intensity in the two respective resonators. In particular, such kind of nonlinearity is attainable in  nonlinear topolectrical circuits by utilizing varactor diodes~\cite{Hadad_2018}.

We describe the state of the system by the column-vector wave function $\ket{\Psi}=\left(\Psi_1,\Psi_2,\dots,\Psi_N \right)^T$, where $\Psi_n$ is the field amplitude at the $n^{\text{th}}$ site. Then the evolution of $\ket{\Psi}$ is captured by the Shr{\"o}dinger equation
\begin{equation}\label{eq:Schr}
    i\df{\ket{\Psi}}{t}=\hat{H}(\Psi)\,\ket{\Psi}\:,
\end{equation}
where the effective Hamiltonian reads
\begin{equation}\label{eq:Hamiltonian}
  \hat{H}(\Psi) =
  \left[ {\begin{array}{ccccc}
    \omega_0 & J_1 & \ddots & 0 & 0\\
    J_1 & \omega_0 & J_2^{(2,3)} & \ddots & 0 \\
    \ddots & J_2^{(3,2)} & \omega_0 & \ddots & \ddots \\
    0 & \ddots & \ddots & \omega_0 & J_1 \\
    0 & 0 & \ddots & J_1 & \omega_0 \\
  \end{array} } \right]\:.
\end{equation}
Here, $\omega_0$ is the eigenfrequency of an isolated resonator. From now on, we set $\omega_0$ to 0, i.e. all frequencies are measured relative to $\omega_0$ level.  The nonlinear coupling is defined by the expression $J_2^{(2n,2n+1)}=J_2+\alpha (|\Psi_{2n}|^2+|\Psi_{2n+1}|^2)$, where $J_2>0$ is a linear part of the intercell coupling and $\alpha$ is the Kerr-type nonlinearity. In the case of topolectrical circuits, it is achieved by inserting the varactor diode between the two $LC$ resonators.

The wave function is normalized such that the value $|\Psi_{2n-1}|^2+|\Psi_{2n}|^2$ represents the energy stored in the $n^{\text{th}}$ dimer. Also, following Ref.~\cite{Hadad_2016}, we define the intensity as $I = \text{max}_n\{\sqrt{|\Psi_{2n-1}|^2+|\Psi_{2n}|^2}\}$.

The linear counterpart of the model Eqs.~\eqref{eq:Schr}-\eqref{eq:Hamiltonian} at $\alpha=0$ exhibits a midgap topological edge state with frequency $\omega_0$ provided the first bond at the termination is weak, $J_1<J_2$.
Although at $\alpha \ne 0$ the problem becomes nonlinear, we can still search for 
the steady-state solutions in the form of the edge-localized states. For this purpose, we adopt the harmonic ansatz for the time-dependent wave function:
\begin{equation}
    \ket{\Psi(t)}=\ket{\Psi}e^{-i\omega t}\:.
\end{equation}
This converts the Schr{\"o}dinger equation~\eqref{eq:Schr} to the nonlinear eigenvalue problem:
\begin{equation}
    \hat{H}(\Psi)\ket{\Psi}=\omega \ket{\Psi}\:.
\end{equation}

If the array is finite and consists of an even number of sites, the edge-localized modes may appear at both edges of the array. Due to the finite length of the array, the edge modes hybridize forming symmetric and anti-symmetric combinations, while their frequencies $\omega$ shift away from the zero value. 

However, the analysis is significantly simplified for the array with the odd number of sites. In such a case, the chiral symmetry of the model ensures that the edge mode residing at one (odd-site) sublattice has exactly zero frequency $\omega=0$. This property allows to simplify the problem further and recast the nonlinear eigenvalue problem into a system of algebraic equations, where the amplitudes at all even nodes are zero $\Psi_{2n} = 0$, while the equations for the odd sites read:
\begin{equation}\label{eq:EdgeModeProfile}
    0 = J_1 \Psi_{2n-1} + J_2 \Psi_{2n-1} +\alpha |\Psi_{2n+1}|^2 \Psi_{2n+1}.
\end{equation}
Thus, the profile of the edge mode in an array with an odd number of sites can be readily computed.

Below, we examine the situation in which the linear couplings satisfy the condition $J_1 > J_2$ and set $\alpha>0$.  Accordingly, in the low-intensity limit, the system [Fig.~\ref{nonlin}(a)] has no edge states at the left edge and supports a localized mode at the right edge. However, when the intensity in the array is increased above the critical value $I_{\text{cr}} = \sqrt{{(J_1-J_2)}/{\alpha}}$, the localization of the edge mode changes from one edge to the other, signalling a self-induced topological transition~\cite{Hadad_2016}.


The profile of the self-induced edge state, calculated for the effective parameters corresponding to those attainable in experiments with nonlinear topolectrical circuits~\cite{Hadad_2018}, is depicted in Fig.~\ref{nonlin}(b). In line with Ref.~\cite{Hadad_2016}, the self-induced topological mode features a non-decreasing tail such that $|\Psi_{n}| \rightarrow I_{\text{cr}}$ for large $n$. 
%
%
This tail seemingly creates a physical paradox, since such a mode in a semi-infinite array stores an infinite amount of energy. To clarify this, we simulate the excitation of the nonlinear edge mode via the left edge as described by the coupled-mode equations
\begin{equation}
    i\frac{\partial}{\partial t}\ket{\Psi} = \hat{H}(\Psi)\ket{\Psi} +\xi \ket{S(t)},
\end{equation}
where $\ket{S(t)} = \left( S(t),0,\dots,0 \right)^T$ is the pump profile localized at the first site. Examining various types of time-dependent excitation $S(t)$ with a spectrum matching the frequency of the edge mode, we observe that the profile of the edge mode builds up gradually, as expected for tight-binding lattices. In the case of the simulated finite lattices, the excitation of the edge mode takes a finite time which increases with the length of the array. Hence, in the limit of an infinite array, the excitation of the edge mode will take an infinitely long time, resolving the paradox mentioned above.



Another interesting aspect of the system is the structure of the nonlinear edge mode at low intensities, slightly lower than the threshold value $I_{\text{cr}}$ [Fig.~\ref{nonlin}(c)]. In this scenario, the intensity at the left edge is {\it lower} than the saturation value $I_{\text{cr}}$, and the deficit of intensity decreases exponentially in the bulk of the array. Such type of localization does not occur in the linear SSH model and becomes possible due to the nonlinearity of the system.

\begin{figure}[b]
  \centering
  \includegraphics[width=1\linewidth]{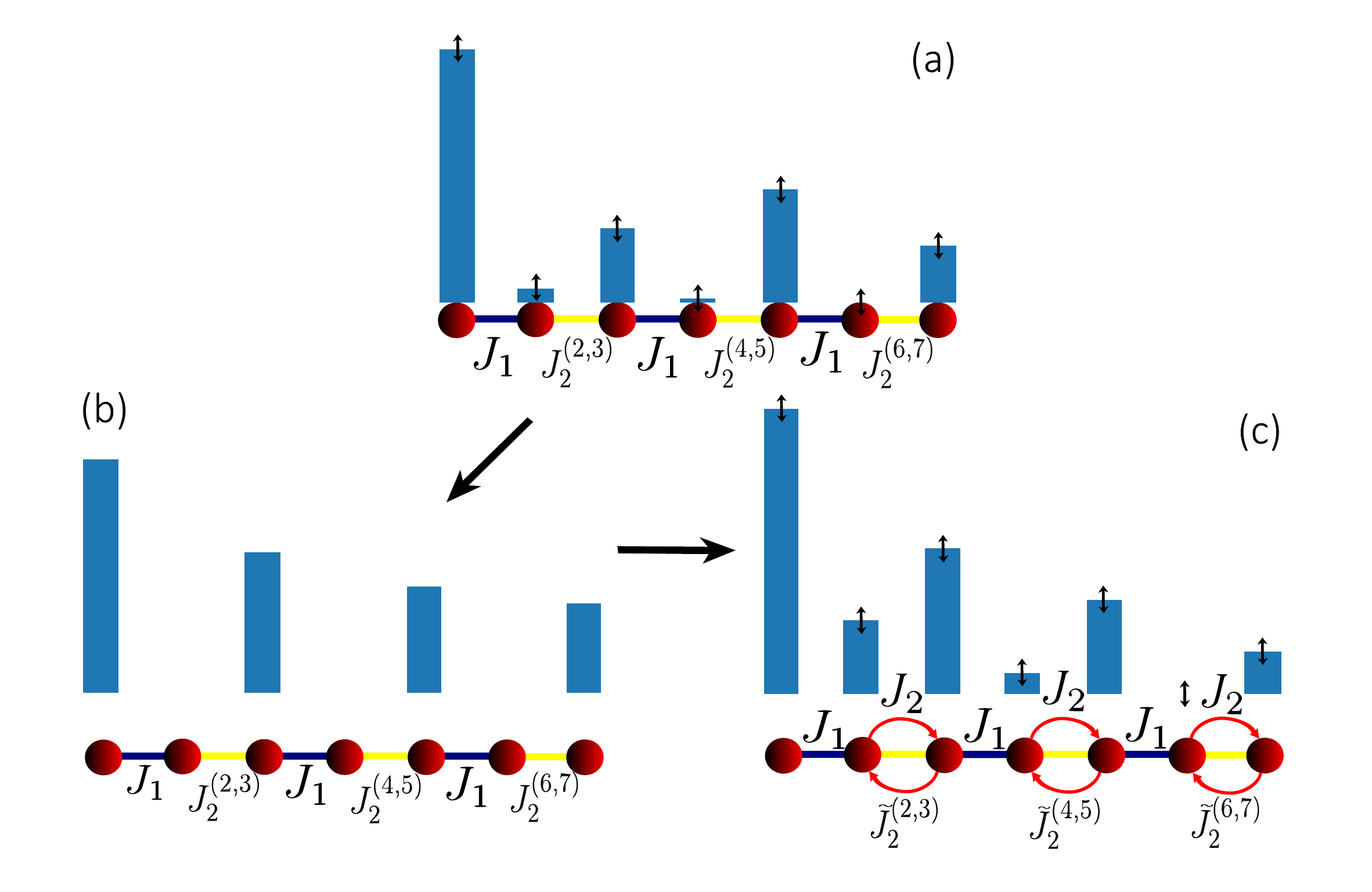}
  \caption{Schematic representation of the relationship between non-Hermitian and nonlinear phenomena. Small perturbation of a nonlinear edge state (a) can be represented as the sum of an undisturbed nonlinear edge state (b) and effectively non-Hermitian excitation in a system where the hopping coefficients from left to right and from right to left are not equal to each other (c).}
  \label{schematic}
\end{figure}

\section{Small perturbations}\label{sec:perturb}

As a next step, we examine small perturbations on top of the self-induced nonlinear topological state $\ket{\Psi_0(t)}$ following the standard linearization procedure. This method is instrumental in analyzing the modulational instability of both the bulk and edge nonlinear steady states; see, for example, Refs.~\cite{Leykam2021,Smolina2023} pertaining to nonlinear topological lattices. We represent the wave function in the form $\ket{\Psi(t)} = \ket{\Psi_0(t)} +\delta \ket{\phi(t)}$, where $\delta$ is a small parameter quantifying the strength of the perturbation, and $\ket{\phi(t)}$ describes the spatio-temporal structure of the perturbation. In the same logic, we expand the Hamiltonian $\hat{H}(\Psi_0+\delta \phi) = \hat{H}(\Psi_0)+\delta \hat{M}(\Psi_0, \phi)+\delta^2\hat{N}(\phi)$ and keep the terms up to the first order in $\delta$. Combining this with Eq.~\eqref{eq:Schr}, we recover 
%
%
\begin{equation}\label{eq:a}
    i\frac{\partial}{\partial t}\ket{\phi} = \hat{H}_0\ket{\phi} +\hat{A}\ket{\phi}\ +e^{-2i\omega t}\hat{A}\ket{\phi^*}\:,
\end{equation}
where $\hat{H}_0 = \hat{H}(\Psi_0)$ and the matrix $\hat{A}$ has a block-diagonal form
\begin{equation}
    \hat{A}=\text{diag}\left(0,\hat{C}_1,\hat{C}_2,\dots \hat{C}_{N-1}\right)\:,
\end{equation}
%
%
while the auxiliary matrices $\hat{C}_n$ are defined as
\begin{equation}
  \hat{C}_n =
  \alpha\left[ {\begin{array}{cc}
  \Psi^{0*}_{2n}\Psi^0_{2n+1}&\Psi^{0*}_{2n+1}\Psi^0_{2n+1}\\
\Psi^{0*}_{2n}\Psi^0_{2n}&\Psi^{0}_{2n}\Psi^{0*}_{2n+1}\\
  \end{array} } \right]
  = 
  \alpha \left[ {\begin{array}{cc}
  0 & |\Psi^{0}_{2n+1}|^2 \\
  0 & 0 \\
  \end{array} } \right],
\end{equation}
blocks $\hat{C}_n$ being simplified for the edge state with $\Psi^0_{2n} = 0$. Since we focus on the perturbations on top of the self-induced topological mode with $\omega=0$, the time-dependent factor $e^{-2i\omega t}$ is equal to $1$ resulting in the equation
\begin{equation}\label{eq:nonh_eq}
    i\frac{\partial}{\partial t}\ket{\phi} = \hat{H}_0\ket{\phi} +2\hat{A}\cdot \text{Re}\{\ket{\phi}\}.
\end{equation}
Matrices $\hat{C}_n$ are clearly non-symmetric and non-Hermitian. Physically, this means that the effective coupling terms in Eq.~\eqref{eq:nonh_eq} are nonreciprocal, a feature that makes the problem analogous to the non-Hermitian SSH array with nonreciprocal couplings~\cite{Yao_2018}. Thus, the treatment of small perturbations in the nonlinear system shown in Fig.\ref{schematic}(a) involves two main ingredients: finding the nonlinear edge state depicted in Fig.\ref{schematic}(b) and solving the linear non-Hermitian problem for small perturbations illustrated in Fig.~\ref{schematic}(c).


\begin{figure}[b]
  \centering
    \includegraphics[width=1\linewidth]{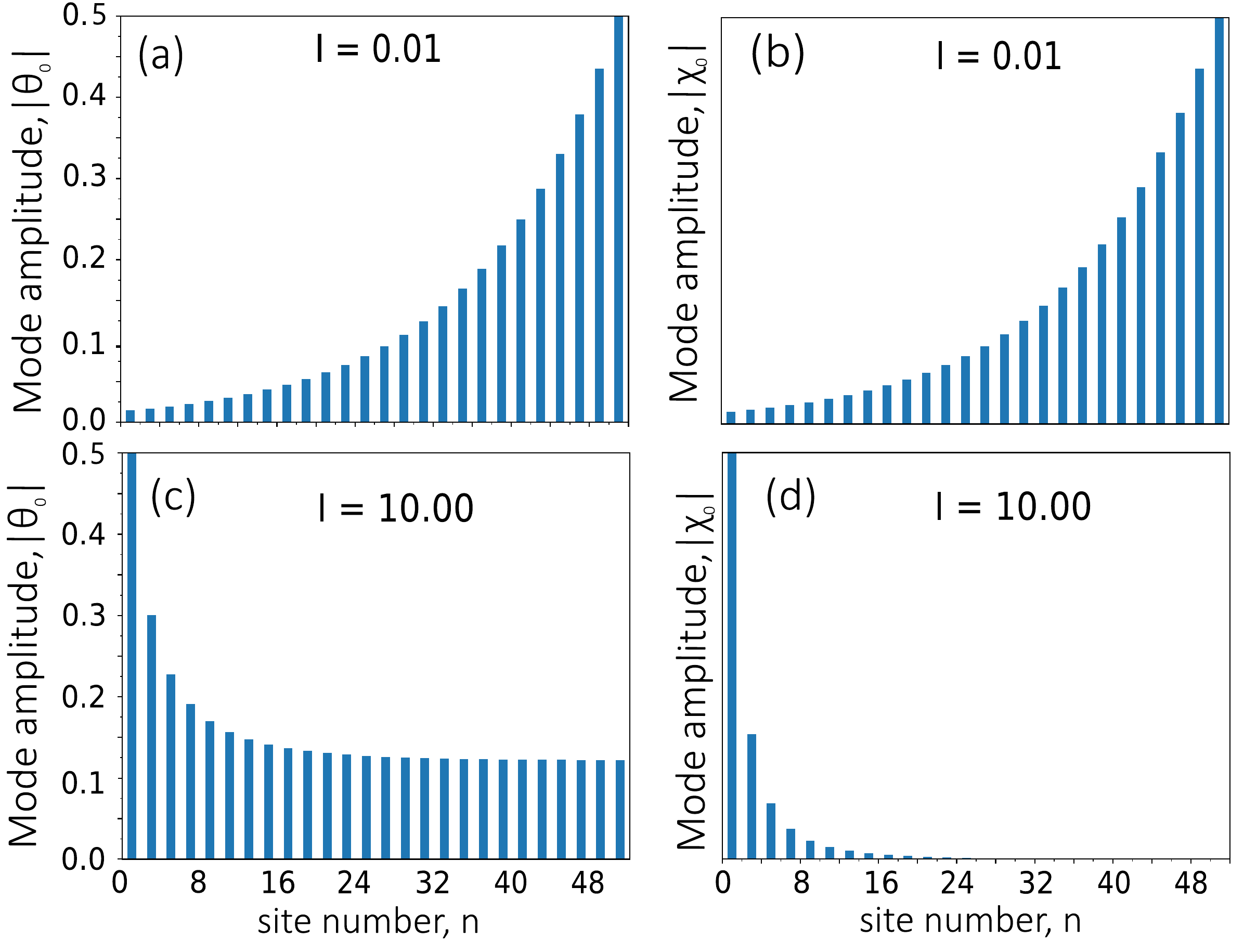}
    \caption{Spatial distribution of the zero-frequency modes describing non-Hermitian excitations on the background of the nonlinear topological state in a finite lattice of $N = 51$ sites. (a,b) Low-intensity regime,  $I=0.01$; (c,d) high-intensity regime, $I=10$. Panels (a,c) and (b,d) show imaginary and real parts of the right eigenvectors, respectively. }
    \label{zero_modes}
\end{figure}

\begin{figure}[b]
    \centering
        \includegraphics[width=1\linewidth]{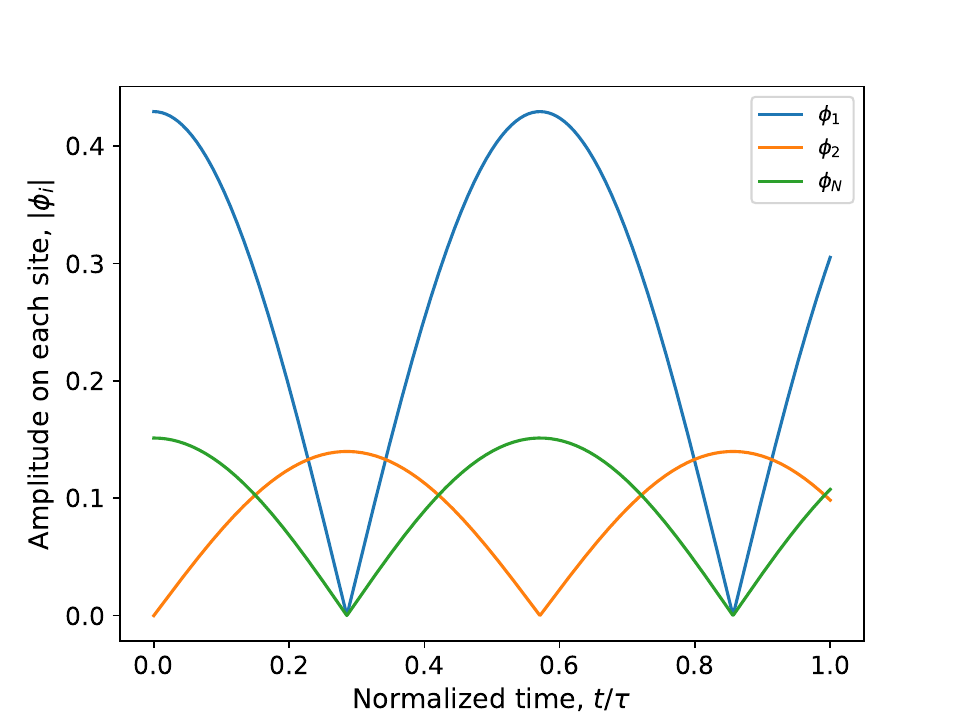}
        \caption{Probability amplitudes $|\phi_n|$ at the sites $1$, $2$ and $N$ for the eigenstate with frequency $\omega = 8.5 \times 10^{-4}$ at intensity level $I/I_{\text{sat}} = 1$ in the chain with $N = 21$ nodes. $\tau = 2\pi / \omega$ is the period of oscillations.
        }
        \label{fig:excitations_behaviour}
\end{figure}

\begin{figure*}
  \centering
  \includegraphics[width=1\linewidth]{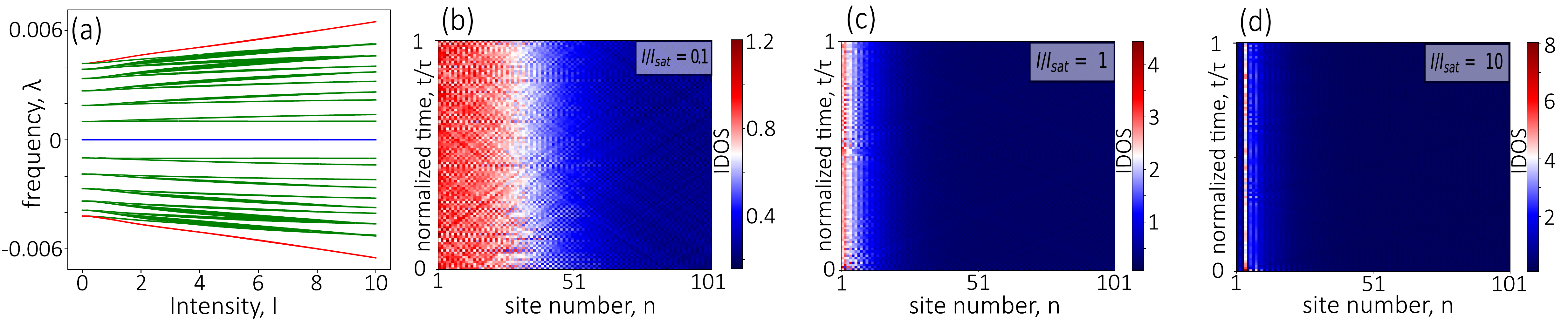}
  \caption{(a) The spectrum of excitations on the background of nonlinear topological edge state as a function of intensity computed for the array consisting of $N=13$ sites. Blue color shows the mode with the zero frequency, red shows the modes with the highest frequencies, and green shows the rest of the modes. IDOS (b,c,d) for different intensity in the array of $N=101$ sites. Parameter $\tau = 2\pi / \omega_{\text{min}}$ is the period of oscillations for mode with the smallest nonzero frequency.}
  \label{spectra}
\end{figure*}

To solve the Eq.~\eqref{eq:nonh_eq}, we separate real and imaginary part of the perturbation  $\ket{\phi}=\ket{\chi}+i\ket{\theta}$. Doubling the dimensionality of the problem, we obtain a system of linear differential equations  
\begin{equation}\label{eq:system}
    \frac{\partial}{\partial t} \begin{pmatrix} \ket{\chi}\\ \ket{\theta} \end{pmatrix} 
    = \begin{pmatrix} \hat{0}_N & \hat{H}_0\\ -\hat{H}_0-2\hat{A} & \hat{0}_N \end{pmatrix}
    \begin{pmatrix} \ket{\chi}\\ \ket{\theta} \end{pmatrix} \:,
\end{equation}
which is straightforward to solve numerically and which yields $2N$ eigenvalues $\mu_n$. Despite the problem is manifestly non-Hermitian, all eigenvalues $\mu_n$  are purely imaginary, which means that the oscillations of $\ket{\phi}$ are non-decaying. We represent these eigenvalues in the form $\mu_n=i\lambda_n$, where $\lambda_n$ are real numbers which appear in pairs $\lambda_{-n}=-\lambda_n$. This demonstrates that the spectrum of perturbations is symmetric with respect to the zero frequency. In addition, the eigenvectors corresponding to $-\lambda_{n}$, $\lambda_n$ eigenvalue pair are complex conjugated: $\textbf{X}_{-n}=\textbf{X}_n^*$.

Zero eigenvalue $\lambda=0$ is doubly degenerate. The respective eigenvectors are:
\begin{equation}\label{eq:ImHerm}
\ket{\chi}=0, \mspace{10mu} \hat{H}_0\ket{\theta}=0, 
\end{equation}
which is the standard zero-energy mode in the Hermitian case and
\begin{equation}\label{eq:ReNonh}
\left(\hat{H}_0+2\,\hat{A}\right)\,\ket{\chi}=0, \mspace{10mu} \ket{\theta}=0, 
\end{equation}
which is the zero-energy mode in the non-Hermitian SSH with nonreciprocal couplings.

By definition, vector $\textbf{X}$ must be real at any moment of time. For $\lambda_n\not=0$ such real-valued solutions are recovered by combining the eigenvectors corresponding to the conjugate eigenvalues:
\begin{equation}\label{eq:vector}
    \tilde{\textbf{X}}_n = \frac{1}{2} \{
    \textbf{X}_n e^{-i\lambda_n t}+{\bf X}_n^* e^{i\lambda_n t} \}\:.
\end{equation}
In terms of Eq.~\eqref{eq:system}, there are $2N$ modes including $2$ with zero frequency and $2N-2$ modes with $\lambda\not=0$. As the latter are necessarily combined together via Eq.~\eqref{eq:vector}, we recover $N+1$ eigenvector $\ket{\phi}$.


Thus, the key distinction of our system from the conventional SSH model is the emergence of {\it two} zero-frequency modes instead of one. The first of them [Eq.~\eqref{eq:ImHerm}] oscillates with $\pi/2$ phase shift with respect to the background edge state and carries no signatures of the non-Hermitian physics satisfying the usual equation Eq.~\eqref{eq:ImHerm}. In contrast, the second mode oscillates in phase with the background mode $\ket{\Psi_0}$, and its profile is modified due to the $\hat{A}$ term in Eq.~\eqref{eq:ReNonh}.

In the low intensity limit, the $\hat{A}$ term is negligibly small and the distinction between the two modes disappears yielding essentially a single zero-energy state known in the Hermitian context [Fig.~\ref{zero_modes}(a,b)]. However, once the intensity is increased, the distinction between the two zero-energy solutions becomes apparent [Fig.~\ref{zero_modes}(c,d)]. While the ``Hermitian'' mode satisfying Eq.~\eqref{eq:ImHerm} shows exponential localization at the edge, its non-Hermitian counterpart Eq.~\eqref{eq:ReNonh} features a non-decreasing tail resembling that of the nonlinear edge state $\ket{\Psi_0}$.

Yet another signature of non-Hermitian physics is provided by the modes with $\lambda\not=0$. Due to their structure [Eq.~\eqref{eq:vector}], the real and imaginary parts of the perturbation $\ket{\phi}$ could oscillate {\it with different amplitudes}. As a result, the probability amplitude at a given site $\phi_n=\sqrt{\chi_n^2+\theta_n^2}$ oscillates in time as illustrated in Fig.~\ref{fig:excitations_behaviour}. 



\section{Mode localization}\label{sec:modsloc}

An important feature of non-Hermitian systems is the non-Hermitian skin effect, which results in the exponential localization of all eigenmodes at the edge of the finite structure~\cite{Bergholtz_2021}. To probe such localization, we use the notion of the integrated density of states $\text{IDOS}(n,t)$ which shows the probability to find a particle at a given site $n$ at a given moment of time $t$ summed over all eigenstates of the system. The energies of those eigenstates are depicted in Fig.~\ref{spectra}(a).

As the localization of the zero-energy modes has been explored above, we include their contribution to IDOS expression with a smaller weight $w=0.2$:
\begin{equation}\label{IDOS}
    \text{IDOS}(n, t) = \sum_{m, \omega \neq 0} |\Psi_n^{(m)}(t)|^2 + w\,\sum_{m, \omega = 0} |\Psi_n^{(m)}(t)|^2 
\end{equation}
The summation is performed over all eigenmodes $m = 1,..., N+1$. We calculate IDOS numerically for the different intensity levels showing the results in Fig.~\ref{spectra}(b-d).

If the intensity is close to zero, the system is similar to the classical SSH. However, even small non-Hermiticity results in the leftward shift of IDOS, suggesting the onset of non-Hermitian skin effect. As the intensity is increased further reaching the threshold level $I_{\text{cr}}$, the leftward shift becomes pronounced having a maximum at the second resonator [Fig.~\ref{spectra}(c)]. Finally, when the intensity exceeds the critical level by an order of magnitude, IDOS becomes strongly localized featuring a maximum around the fourth site.


While the probability distribution clearly shifts to the left, this behavior is not a pure non-Hermitian skin effect, as not all of the modes are exponentially localized at the left edge. This is rooted to the phase-dependent nature of non-Hermiticity in our system: the mode experiences non-Hermitian corrections depending on the phase shift between the background mode and the respective small perturbation.


\section{Discussion and outlook}\label{sec:Disc}

In summary, we have analysed small perturbations on the background of the self-induced edge state in the nonlinear SSH model. Even though the original model is Hermitian, the behavior of the perturbations is captured by the non-Hermitian effective Hamiltonian exhibiting several counter-intuitive features.

First, the couplings in the effective model are manifestly non-reciprocal implying different tunneling amplitudes in the opposite directions. This opens an easy access to the experimental realization of SSH models with nonreciprocal couplings which became a focus of intense theoretical investigations~\cite{Kunst_2018,Yao_2018} and have recently been realized experimentally in complicated systems involving active elements~\cite{Helbig2020} or temporal modulation~\cite{Weidemann2020}.

Second, even stationary solutions for the perturbation $\ket{\phi}$ exhibit oscillating probability amplitudes at the sites of the array pointing towards persistent currents in the system, absent in the Hermitian scenario.

Finally, non-Hermitian nature of the model results in the overall shift of the probability distribution towards the edge of the array, resulting in a modification of the well-celebrated non-Hermitian skin effect.

We believe that these results delineate a route to access non-Hermitian phenomena via small perturbations of nonlinear steady waves and to probe the associated effects experimentally.



\section*{Acknowledgments}

Theoretical models were supported by the Russian Science Foundation (Grant No.~23-72-10026). Numerical simulations were supported by Priority 2030 Federal Academic Leadership Program.  M.A.G. acknowledges partial support from the Foundation for the Advancement of Theoretical Physics and Mathematics ``Basis''. D.A.S. acknowledges support from the Australian Research Council (FT230100058).

\bibliography{main}

\end{document}